\documentclass[aps,prl,twocolumn,groupedaddress]{revtex4}

\usepackage[dvips]{graphicx,color} 
\usepackage{array,hhline,dcolumn} 
\usepackage{rotating} 

\newcommand{\gtwid}{\mathrel{\raise.3ex\hbox{$>$\kern-.75em\lower1ex\hbox{$\sim$}}}}
\newcommand{\ltwid}{\mathrel{\raise.3ex\hbox{$<$\kern-.75em\lower1ex\hbox{$\sim$}}}}

\begin{document}
%

\title{A Combined $\nu_\mu \rightarrow \nu_e$ \& $\bar \nu_\mu \rightarrow \bar \nu_e$ Oscillation Analysis of the MiniBooNE Excesses}

\author{
        A.~A. Aguilar-Arevalo$^{12}$, 
        B.~C.~Brown$^{6}$, L.~Bugel$^{11}$,
	G.~Cheng$^{5}$, E.~D.~Church$^{15}$, J.~M.~Conrad$^{11}$,
	R.~Dharmapalan$^{1}$, 
	Z.~Djurcic$^{2}$, D.~A.~Finley$^{6}$, R.~Ford$^{6}$,
        F.~G.~Garcia$^{6}$, G.~T.~Garvey$^{9}$, 
        J.~Grange$^{7}$,
        W.~Huelsnitz$^{9}$, C.~Ignarra$^{11}$, R.~Imlay$^{10}$,
        R.~A. ~Johnson$^{3}$, G.~Karagiorgi$^{5}$, T.~Katori$^{11}$,
        T.~Kobilarcik$^{6}$, 
        W.~C.~Louis$^{9}$, C.~Mariani$^{5}$, W.~Marsh$^{6}$,
        G.~B.~Mills$^{9}$,
	J.~Mirabal$^{9}$,
        C.~D.~Moore$^{6}$, J.~Mousseau$^{7}$, 
        P.~Nienaber$^{14}$, 
        B.~Osmanov$^{7}$, Z.~Pavlovic$^{9}$, D.~Perevalov$^{6}$,
        C.~C.~Polly$^{6}$, H.~Ray$^{7}$, B.~P.~Roe$^{13}$,
        A.~D.~Russell$^{6}$, 
	M.~H.~Shaevitz$^{5}$, 
        J.~Spitz$^{11}$, I.~Stancu$^{1}$, 
        R.~Tayloe$^{8}$, R.~G.~Van~de~Water$^{9}$, 
        D.~H.~White$^{9}$, D.~A.~Wickremasinghe$^{3}$, G.~P.~Zeller$^{6}$,
        E.~D.~Zimmerman$^{4}$ \\
\smallskip
(The MiniBooNE Collaboration)
\smallskip
}
\smallskip
\smallskip
\affiliation{
$^1$University of Alabama; Tuscaloosa, AL 35487 \\
$^2$Argonne National Laboratory; Argonne, IL 60439 \\
$^3$University of Cincinnati; Cincinnati, OH 45221\\
$^4$University of Colorado; Boulder, CO 80309 \\
$^5$Columbia University; New York, NY 10027 \\
$^6$Fermi National Accelerator Laboratory; Batavia, IL 60510 \\
$^7$University of Florida; Gainesville, FL 32611 \\
$^8$Indiana University; Bloomington, IN 47405 \\
$^9$Los Alamos National Laboratory; Los Alamos, NM 87545 \\
$^{10}$Louisiana State University; Baton Rouge, LA 70803 \\
$^{11}$Massachusetts Institute of Technology; Cambridge, MA 02139 \\
$^{12}$Instituto de Ciencias Nucleares, Universidad Nacional Aut\'onoma de M\'exico, D.F. 04510, M\'exico \\
$^{13}$University of Michigan; Ann Arbor, MI 48109 \\
$^{14}$Saint Mary's University of Minnesota; Winona, MN 55987 \\
$^{15}$Yale University; New Haven, CT 06520\\
}

\date{\today}

\begin{abstract}
The MiniBooNE experiment at Fermilab reports results from an 
analysis of the combined $\nu_e$ and $\bar \nu_e$ appearance data
from $6.46 \times 10^{20}$ protons
on target in neutrino mode and $11.27 \times 10^{20}$ protons
on target in antineutrino mode. A total excess
of $240.3 \pm 34.5 \pm 52.6$ events ($3.8 \sigma$) is observed from
combining the two data sets in the
energy range $200<E_\nu^{QE}<1250$~MeV.
In a combined fit for CP-conserving $\nu_\mu \rightarrow \nu_e$ and
$\bar{\nu}_{\mu}\rightarrow\bar{\nu}_e$ oscillations via a 
two-neutrino model,
the background-only fit has a $\chi^2$-probability of 0.03\% relative to the best
oscillation fit. The data are consistent with neutrino 
oscillations in the $0.01 < \Delta m^2 < 1.0$ eV$^2$ range and with the
evidence for antineutrino oscillations from the Liquid Scintillator 
Neutrino Detector (LSND).
\end{abstract}

\pacs{14.60.Lm, 14.60.Pq, 14.60.St}

\keywords{Suggested keywords}
\maketitle


There is growing evidence for short-baseline neutrino anomalies occuring at an $L/E_\nu \sim 1$ m/MeV,
where $E_\nu$ is the neutrino energy and $L$ is the distance that the neutrino travelled before detection.
These anomalies include the excess of events observed by the LSND \cite{lsnd} and MiniBooNE 
\cite{mb_osc,mb_lowe,mb_osc_anti}
experiments and the deficit of events observed by reactor \cite{reactor}
and radioactive-source experiments \cite{radioactive}.
There have been several attempts to interpret these anomalies in terms of 3+N 
neutrino oscillation models 
involving three active neutrinos and N additional sterile neutrinos 
\cite{sorel,karagiorgi,giunti,kopp,white_paper,3+2}. 
(Other more exotic explanations include, for example, Lorentz violation 
\cite{lorentz} and sterile neutrino decay \cite{sterile_decay}.)
This paper presents a combined oscillation analysis
of the MiniBooNE $\nu_e$ and $\bar \nu_e$ appearance data,
corresponding to $6.46 \times 10^{20}$ protons
on target (POT) in neutrino mode \cite{mb_lowe} and $11.27 \times 10^{20}$ POT
in antineutrino mode, which is approximately twice the antineutrino
data reported previously \cite{mb_osc_anti}.

This analysis fits both $\nu_\mu \rightarrow \nu_e$ and $\bar \nu_\mu \rightarrow \bar \nu_e$ oscillations
with the same oscillation model
over the full neutrino energy range $200<E_\nu^{QE}<3000$~MeV, where $E_\nu^{QE}$ is the reconstructed neutrino
energy assuming quasielastic scattering kinematics \cite{MBCCQE}. 
The neutrino oscillation energy region is defined to be $200<E_\nu^{QE}<1250$~MeV, which is where
an LSND-like signal (same $L/E_\nu$) is expected. 
Combining neutrino and antineutrino data over the full energy range has the advantage of decreasing statistical and
systematic errors.
The analysis assumes no significant $\nu_\mu$, $\bar{\nu}_{\mu}$, $\nu_e$, 
or $\bar \nu_e$ disappearance. This simplification
may change the fitted $\nu_\mu \rightarrow \nu_e$ and $\bar \nu_\mu \rightarrow \bar \nu_e$ 
appearance probabilities by up to $\sim 20\%$. Furthermore, it has been suggested that
nuclear effects associated with neutrino interactions on carbon can affect the reconstruction of the
neutrino energy and the determination of the neutrino oscillation parameters \cite{nuclear_effects}.
These effects are not fully accounted for in the analysis
and may affect somewhat the 
oscillation fit parameters discussed below.

The neutrino (antineutrino) flux is produced by 8 GeV protons from the Fermilab Booster
interacting on a beryllium target inside a magnetic focusing horn set at positive 
(negative) polarity. In neutrino (antineutrino) mode, positively 
(negatively) charged mesons 
produced in p-Be interactions are focused in the forward direction 
and subsequently decay primarily into $\nu_\mu$ ($\bar{\nu}_{\mu}$). The flux of
neutrinos and antineutrinos of all flavors is simulated
using information from external measurements \cite{mb_flux}.
In neutrino mode, the $\nu_\mu$, $\bar \nu_\mu$, $\nu_e$, and $\bar \nu_e$ flux
contributions at the detector are 93.5\%, 5.9\%, 0.5\%, and 0.1\%, respectively.
In antineutrino mode, the $\bar \nu_\mu$, $\nu_\mu$, $\bar \nu_e$, and $\nu_e$ flux
contributions at the detector are 83.7\%, 15.7\%, 0.4\%, and 0.2\%, respectively.
The $\nu_\mu$ and $\bar{\nu}_{\mu}$ fluxes peak at approximately 600 MeV and 400 MeV, respectively. 

The MiniBooNE detector is described in detail in reference \cite{mb_detector}. 
The detector is located 541 m from the beryllium target and consists of
a 40-foot diameter sphere filled with 806 tons of pure mineral oil (CH$_{2}$). Neutrino interactions in the 
detector produce charged particles (electrons, muons, protons, pions, and kaons) which in turn produce scintillation and Cherenkov light 
detected by the 1520 8-inch photomultiplier tubes (PMTs) that line the interior of the detector and
an optically isolated outer veto region. Event reconstruction and particle identification are derived from
the hit PMT charge and time information.

The signature of $\nu_\mu \rightarrow \nu_e$ and
$\bar \nu_\mu \rightarrow \bar \nu_e$ oscillations 
is an excess of $\nu_e$ and $\bar \nu_e$-induced charged-current quasi-elastic (CCQE) 
events. Reconstruction \cite{mb_recon} and selection requirements of these 
events are almost identical to those from previous 
analyses \cite{mb_lowe,mb_osc_anti} with an average reconstruction efficiency 
of $\sim 10-15\%$ for events generated over the entire volume of the
detector. Recent improvements to the analysis include a
better determination of the intrinsic $\nu_e$ background from $K^+$ decay 
through the measurement of high-energy neutrino events in the SciBooNE experiment \cite{sciboone_kaon},
a combined error matrix for neutrino and antineutrino data
with correlated and uncorrelated errors,
a better determination of neutral-current $\pi^0$ and external event background in antineutrino
mode due to the increase in statistics of the antineutrino mode data sample, and the use of a likelihood fit with
frequentist corrections from fake data studies for both the neutrino-mode and antineutrino-mode data.
The detector cannot distinguish
between neutrino and antineutrino interactions 
on an event-by-event basis. However, the fraction of CCQE events in antineutrino (neutrino) mode that
are due to wrong-sign neutrino (antineutrino) events was determined from the angular 
distributions of muons created in CCQE interactions and
by measuring charged-current single $\pi^+$ events \cite{wrong_sign}.

\begin{table}[t]
\vspace{-0.1in}
\caption{\label{signal_bkgd} \em The expected (unconstrained) number of events
for the $200<E_\nu^{QE}<1250$~MeV neutrino oscillation 
energy range from all of the backgrounds in the $\nu_e$ and $\bar{\nu}_e$ 
appearance analysis and for the LSND expectation of 0.26\% oscillation probability averaged over neutrino energy
for both neutrino mode and antineutrino mode. 
}
\small
\begin{ruledtabular}
\begin{tabular}{ccc}
Process&Neutrino Mode&Antineutrino Mode \\
\hline
$\nu_\mu$ \& $\bar \nu_\mu$ CCQE & 37.1 & 12.9 \\
NC $\pi^0$ & 252.3 & 112.3 \\
NC $\Delta \rightarrow N \gamma$ & 86.8 & 34.7 \\
External Events & 35.3 & 15.3 \\
Other $\nu_\mu$ \& $\bar \nu_\mu$ & 45.1 & 22.3 \\
\hline
$\nu_e$ \& $\bar \nu_e$ from $\mu^{\pm}$ Decay & 214.0 & 91.4 \\
$\nu_e$ \& $\bar \nu_e$ from $K^{\pm}$ Decay & 96.7 & 51.2 \\
$\nu_e$ \& $\bar \nu_e$ from $K^0_L$ Decay & 27.4 & 51.4 \\
Other $\nu_e$ \& $\bar \nu_e$ & 3.0 & 6.7 \\
\hline
Total Background &797.7&398.2 \\
\hline
0.26\% $\bar{\nu}_{\mu}\rightarrow\bar{\nu}_e$ & 233.0 & 100.0 \\
\end{tabular}
\vspace{-0.2in}
\end{ruledtabular}
\normalsize
\end{table}

The predicted $\nu_e$ and $\bar{\nu}_e$ CCQE background events for 
the neutrino oscillation energy range $200<E_\nu^{QE}<1250$~MeV 
are shown in Table \ref{signal_bkgd} for both neutrino mode and antineutrino mode. 
The predicted backgrounds to the $\nu_e$ and $\bar{\nu}_e$ CCQE sample are constrained by 
measurements in MiniBooNE 
and include neutral current (NC) $\pi^{0}$ events \cite{mb_pi0}
with photonuclear interactions, 
$\Delta\rightarrow N\gamma$ radiative decays \cite{hill_zhang}, and 
neutrino interactions external to the detector. 
Other backgrounds from mis-identified $\nu_{\mu}$ or $\bar{\nu}_{\mu}$ 
\cite{mb_numuccqe,mb_numuccpi} and from
intrinsic $\nu_e$ and $\bar{\nu}_e$ events from the $\pi\rightarrow\mu$ decay chain
are constrained and obtain their normalizations from the $\nu_\mu$ and $\bar{\nu}_{\mu}$ CCQE data samples,
which consist of 115,467 (50,456) events in neutrino (antineutrino) mode in the $200<E_\nu^{QE}<1900$ MeV
energy range.

Systematic uncertainties are determined by considering the predicted
effects on the $\nu_\mu$, $\bar{\nu}_{\mu}$, $\nu_e$, and $\bar{\nu}_e$ CCQE rate 
from variations of actual parameters.
These include uncertainties in the neutrino and antineutrino flux estimates, 
uncertainties in neutrino cross sections, most of which are determined by in
situ cross-section 
measurements at MiniBooNE, and uncertainties in 
detector modeling and reconstruction. 
A covariance matrix in bins of $E^{QE}_{\nu}$ is constructed 
by considering the variation from each source of systematic uncertainty on the $\nu_e$ and $\bar{\nu}_e$ CCQE signal, background, and 
$\nu_\mu$ and $\bar{\nu}_{\mu}$ CCQE prediction as a function of $E_{\nu}^{QE}$.
This matrix includes correlations between any of the $\nu_e$ and $\bar{\nu}_e$ CCQE signal and background and 
$\nu_\mu$ and $\bar{\nu}_{\mu}$ CCQE samples, and is used in the $\chi^2$ calculation of the oscillation fit.

\begin{table}[b]
\vspace{-0.2in}
\caption{\label{signal_bkgd3} \em The number of data, fitted (constrained)
background, and excess events in the $\nu_e$ and $\bar{\nu}_e$ analyses for
neutrino mode, antineutrino mode, and combined 
in the neutrino oscillation energy range $200<E_\nu^{QE}<1250$~MeV. The uncertainties include both statistical and constrained 
systematic errors. All known systematic errors are included
in the systematic error estimate.}
\begin{ruledtabular}
\begin{tabular}{cccc}
Mode&Data&Background&Excess \\
\hline
Neutrino Mode&952&$790.0 \pm 28.1 \pm 38.7$&$162.0 \pm 47.8$ \\
Antineutrino Mode&478&$399.6 \pm 20.0 \pm 20.3$&$78.4 \pm 28.5$ \\
Combined&1430&$1189.7 \pm 34.5 \pm 52.6$&$240.3 \pm 62.9$ \\
\end{tabular}
\end{ruledtabular}
\end{table}

\begin{figure}[tbp]
\vspace{-0.1in}\centerline{\includegraphics[angle=0, width=9.0cm]{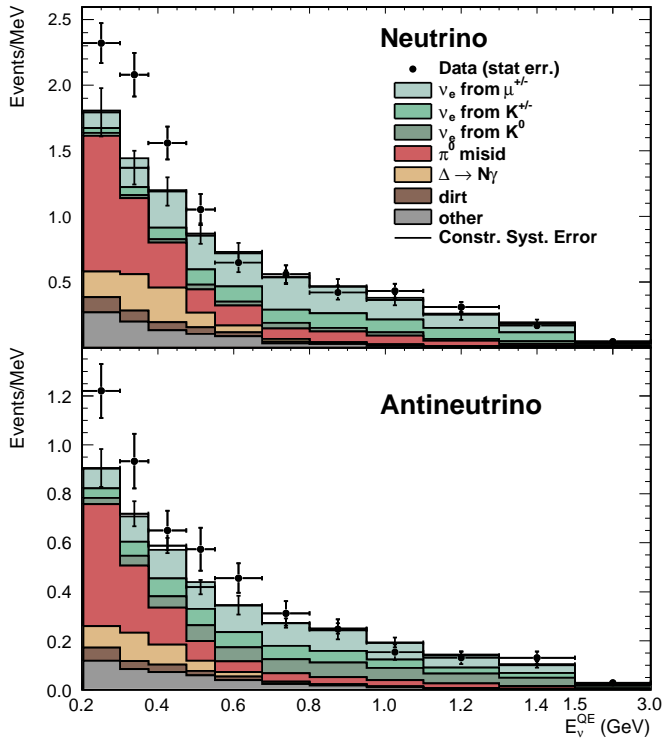}}
\vspace{0.1in}
\caption{The neutrino mode (top) and antineutrino mode (bottom) 
$E_\nu^{QE}$ distributions 
for ${\nu}_e$ CCQE data (points with statistical errors) and background (histogram with systematic errors).} 
\label{excessnat}
\vspace{-0.2in}
\end{figure}

\begin{figure}[tbp]
\vspace{-0.1in}\centerline{\includegraphics[angle=0, width=9.0cm]{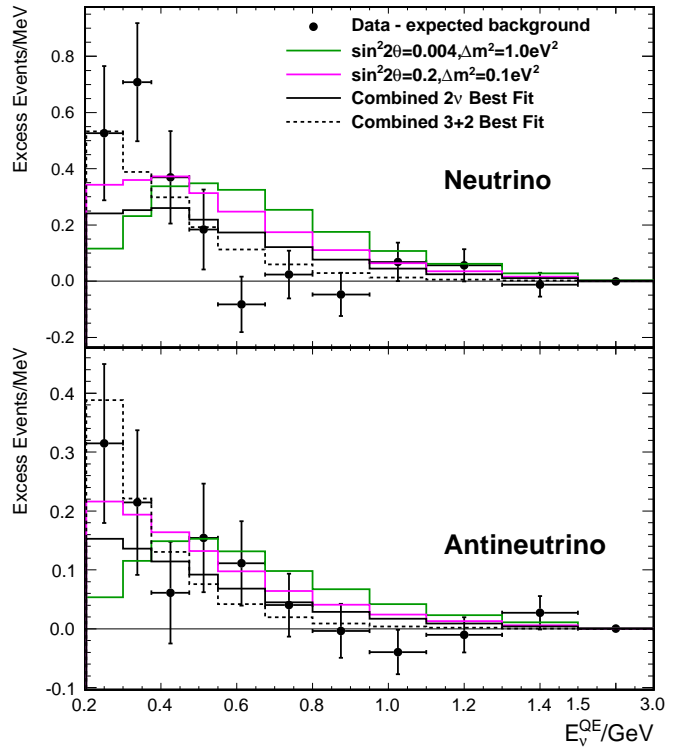}}
\vspace{0.1in}
\caption{The neutrino mode (top) and antineutrino mode (bottom)
event excesses as a function of $E_\nu^{QE}$. Also shown are the 
expectations from the best two-neutrino and 3+2 joint oscillation fits 
with $200<E_\nu^{QE}<3000$~MeV and from two reference values in the LSND
allowed region.
All known systematic errors are included in the systematic error estimate.}
\label{excessnab}
\vspace{-0.2in}
\end{figure}

Fig. \ref{excessnat} shows the $E_\nu^{QE}$ distribution for 
$\nu_e$ and $\bar{\nu}_e$ CCQE data and background in neutrino and
antineutrino mode over the full available energy range. 
Each bin of reconstructed $E_\nu^{QE}$
corresponds to a distribution of ``true'' generated neutrino energies,
which can overlap adjacent bins.
In neutrino (antineutrino) mode, a total of 952 (478) events pass 
the ${\nu}_e$ event selection requirements with $200<E_\nu^{QE}<1250$~MeV, 
compared to an expectation of $790.0 \pm 28.1 \pm 38.7$ ($399.6 \pm 20.0 \pm 20.3$) events, where the 
first error is statistical and the second error is systematic. 
This corresponds to a neutrino (antineutrino) excess of $162.0 \pm 47.8$ ($78.4 \pm 28.5$) events. Combining
the data in neutrino mode and antineutrino mode, the total excess is $240.3 \pm 62.9$ 
events. Fig.~\ref{excessnab} shows the event excesses 
as a function of $E_\nu^{QE}$ in both neutrino and antineutrino modes.
The number of data, fitted background, and excess events for neutrino mode,
antineutrino mode, and combined are summarized in Table \ref{signal_bkgd3}.

Many checks have been performed on the data, including 
beam and detector stability checks that show that the neutrino event 
rates are stable to $<2\%$ and
that the detector energy response is stable to $<1\%$ over the entire run. 
In addition, the fractions of neutrino
and antineutrino events are stable over energy 
and time, and the inferred external event rate corrections are similar
in both neutrino and antineutrino modes. 

\begin{figure}[tbp]
\vspace{-0.1in}\centerline{\includegraphics[angle=0, width=9.0cm]{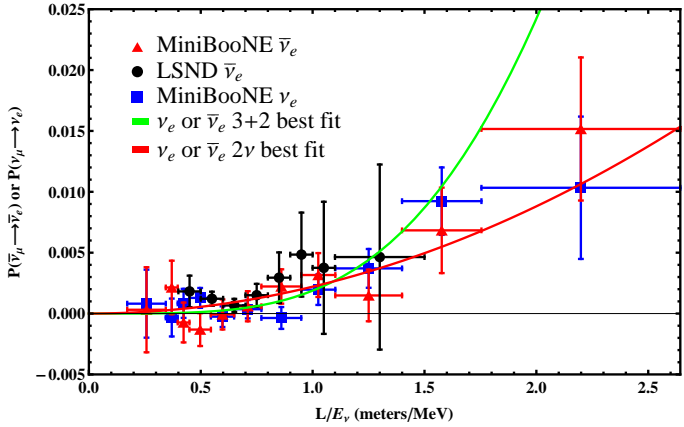}}
\vspace{0.1in}
\caption{The oscillation
probability as a function of $L/E_\nu^{QE}$ for $\nu_\mu \rightarrow
\nu_e$ and $\bar \nu_\mu \rightarrow
\bar \nu_e$ candidate events from MiniBooNE and $\bar \nu_\mu \rightarrow
\bar \nu_e$ candidate events from LSND. The data points include
both statistical and systematic errors. Also shown are the 
oscillation probabilities from the two-neutrino and 3+2 joint oscillation fits.}
\label{L_E}
\vspace{-0.2in}
\end{figure}

\begin{figure}[tbp]
\centerline{\includegraphics[width=9.0cm]{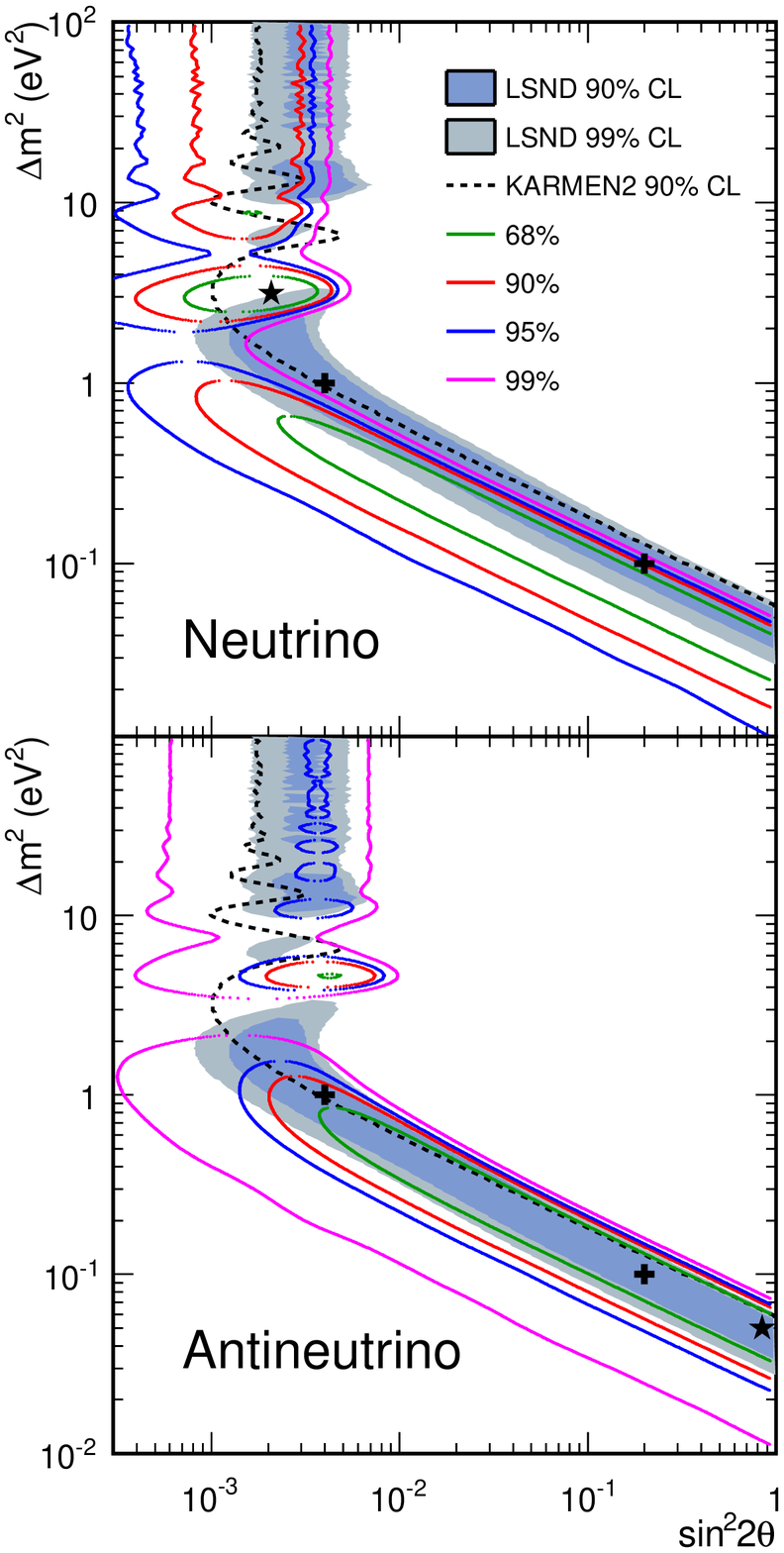}}
\caption{MiniBooNE allowed regions in neutrino mode (top) and
antineutrino mode (bottom) for events with
$E^{QE}_{\nu} > 200$ MeV within a two-neutrino ${\nu}_{\mu}\rightarrow{\nu}_e$ and
$\bar{\nu}_{\mu}\rightarrow\bar{\nu}_e$ oscillation model. 
Also shown is the $\bar{\nu}_{\mu}\rightarrow\bar{\nu}_e$
limit from the KARMEN experiment \cite{karmen}.
The shaded areas show the 90\% and 99\% C.L. LSND 
$\bar{\nu}_{\mu}\rightarrow\bar{\nu}_e$ allowed 
regions. The black stars show the best fit points, while the crosses
show LSND reference values.}\label{limitab}
\vspace{-0.2in}
\end{figure}

\begin{figure}[tbp]
\centerline{\includegraphics[width=9.0cm]{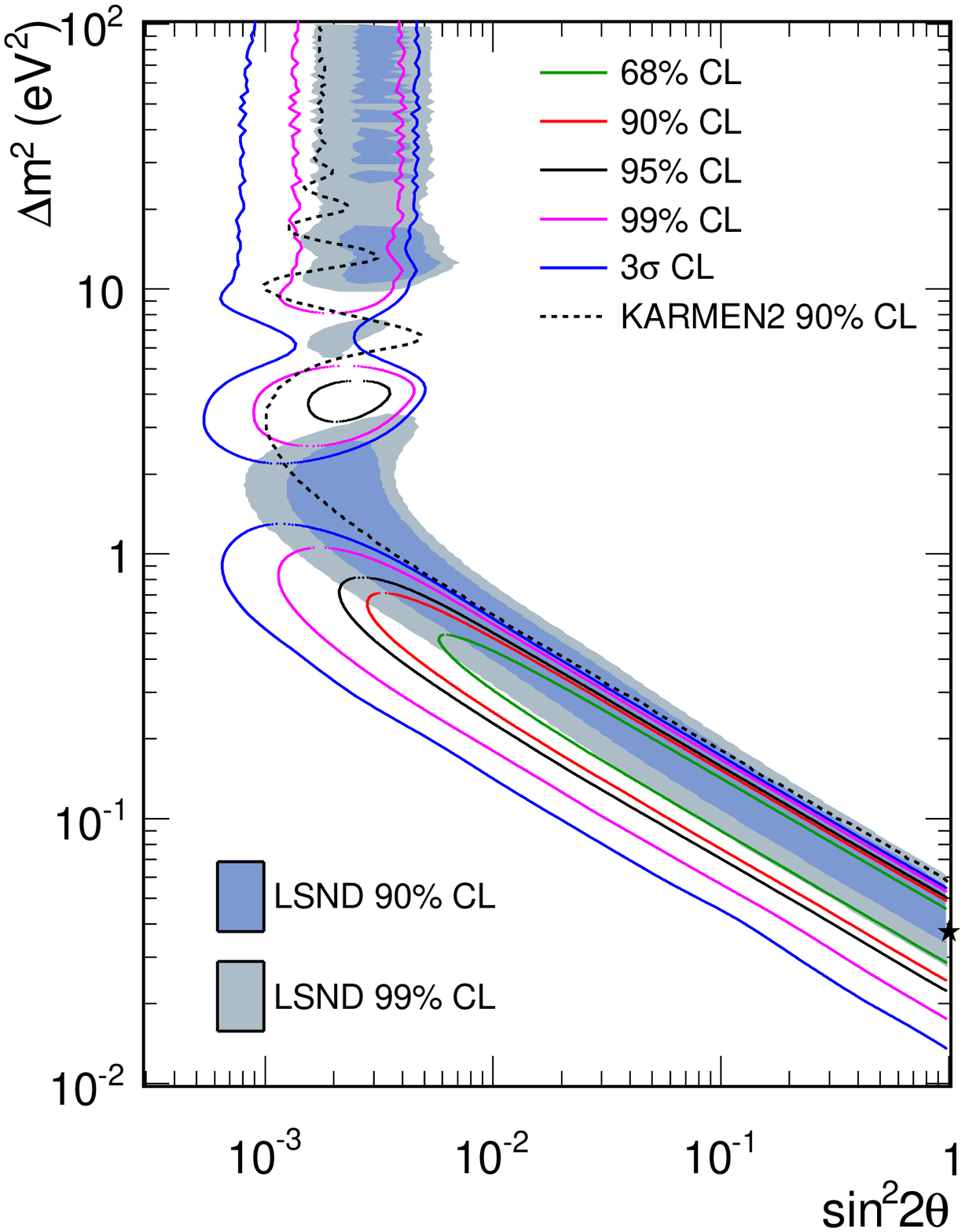}}
\caption{MiniBooNE allowed regions in combined neutrino and antineutrino mode for events with
$200<E^{QE}_{\nu}< 3000$~MeV within a two-neutrino ${\nu}_{\mu}\rightarrow{\nu}_e$ and
$\bar{\nu}_{\mu}\rightarrow\bar{\nu}_e$ oscillation model. 
Also shown is the $\bar{\nu}_{\mu}\rightarrow\bar{\nu}_e$
limit from the KARMEN experiment \cite{karmen}.
The shaded areas show the 90\% and 99\% C.L. LSND 
$\bar{\nu}_{\mu}\rightarrow\bar{\nu}_e$ allowed 
regions. The black star shows the best fit point.}\label{limit}
\vspace{-0.2in}
\end{figure}

A comparison between the MiniBooNE and LSND antineutrino data sets
is given in Fig. \ref{L_E}, which shows the oscillation
probability as a function of $L/E_\nu$ for $\nu_\mu \rightarrow
\nu_e$ and  $\bar \nu_\mu \rightarrow
\bar \nu_e$ candidate events in the $L/E_\nu$ range where
MiniBooNE and LSND overlap. The data used for LSND and MiniBooNE correspond
to $20<E_\nu<60$ MeV and $200<E_\nu^{QE}<3000$~MeV, respectively.
The oscillation probability is defined as the
event excess divided by the number of events expected for 100\% 
$\nu_\mu \rightarrow \nu_e$ and
$\bar \nu_\mu \rightarrow \bar \nu_e$ transmutation in each bin, while $L$
is the distance travelled by the neutrino or antineutrino from the
mean neutrino production point to the detector and
$E_\nu$ is the reconstructed neutrino or antineutrino energy. 
The largest oscillation probabilities from both LSND and MiniBooNE 
occur at $L/E_\nu \ge 1$ m/MeV.

The MiniBooNE data are next fit to a two-neutrino oscillation model, where the
probability, $P$, of $\nu_\mu \rightarrow \nu_e$ and $\bar \nu_\mu \rightarrow
\bar \nu_e$ oscillations is given by $P=\sin^22\theta \sin^2(1.27 \Delta m^2 
L/E_\nu)$, $\sin^22\theta = 4|U_{e4}|^2|U_{\mu4}|^2$, and $\Delta m^2= \Delta m^2_{41} = m^2_4-m^2_1$. 
The oscillation parameters are extracted from a combined fit to the 
$\nu_e$, $\bar \nu_e$, $\nu_\mu$, and $\bar \nu_\mu$ CCQE event distributions.
The fit assumes CP conservation with
the same oscillation probability for neutrinos and antineutrinos, 
including both right-sign and wrong-sign neutrinos, and
no significant $\nu_\mu$, $\bar{\nu}_{\mu}$, $\nu_e$,
or $\bar \nu_e$ disappearance. 
Using a likelihood-ratio technique \cite{mb_osc_anti},
the best oscillation fit for $200<E_\nu^{QE}<3000$~MeV occurs at
($\Delta m^2$, $\sin^22\theta$) $=$ (0.037 eV$^2$, 1.00). 
The $\chi^2/ndf$ for the best-fit point in the
neutrino oscillation energy range of $200<E_\nu^{QE}<1250$~MeV is 
24.7/15.6, corresponding to a probability of 6.7\%.
The probability of the background-only fit relative to the best
oscillation fit is 0.03\%.
Fig.~\ref{limitab} shows the MiniBooNE closed contours for 
$\nu_e$ and $\bar \nu_e$ appearance oscillations in neutrino mode and
antineutrino mode separately in the 
$200<E_\nu^{QE}<3000$~MeV energy range, where a two-neutrino 
oscillation model is
assumed and where frequentist studies were performed
to determine the confidence level (C.L.) 
regions. The
separate best fits for neutrino mode and antineutrino mode are
at ($\Delta m^2$, $\sin^22\theta$) values of (3.14 eV$^2$, 0.002) and (0.05
eV$^2$, 0.842). In the neutrino oscillation energy range of 
$200<E_\nu^{QE}<1250$~MeV, the $\chi^2/ndf$ for the best-fit points 
in neutrino mode and antineutrino mode are 
13.2/6.8 and 4.8/6.9 with probabilities of 6.1\% and 67.5\%, respectively.
The background-only fit has a $\chi^2$-probability of 
1.6\% and 0.5\% relative to the
best oscillation fits for neutrino and antineutrino, respectively.
Fig.~\ref{limit} shows the closed contours for the combined fit.
The allowed regions for $\Delta m^2 < 1$ eV$^2$ are in agreement with the LSND allowed region \cite{lsnd}
and consistent with the limits from the KARMEN experiment \cite{karmen}.
Fig.~\ref{excessnab} shows the expectations from both 
the best two-neutrino joint oscillation fit and from a 3+2 joint 
oscillation fit 
as a function of $E_\nu^{QE}$ in both neutrino and antineutrino modes. 
The best-fit parameters from the 3+2 oscillation fit are
$\Delta m^2_{41} = 0.082$ eV$^2$, $\Delta m^2_{51} = 0.476$ eV$^2$,
$|U_{e4}|^2|U_{\mu 4}|^2=0.1844$, $|U_{e5}|^2|U_{\mu 5}|^2=0.00547$, and $\phi = 1.0005 \pi$.
The 3+2 fit has 
three more parameters than the two-neutrino fit \cite{3+2}
and will be discussed in a future publication.

In summary, the MiniBooNE experiment 
observes a total excess of $240.3 \pm 62.9$ $\nu_e$ and $\bar{\nu}_e$ events ($3.8 \sigma$) in the 
neutrino oscillation energy range $200<E_\nu^{QE}<1250$~MeV.
The allowed regions from a two-neutrino fit to the data, shown in Fig. \ref{limit}, 
are consistent with ${\nu}_{\mu}\rightarrow{\nu}_e$ and
$\bar{\nu}_{\mu}\rightarrow\bar{\nu}_e$ oscillations 
in the 0.01 to 1 eV$^2$ $\Delta m^2$ range 
and consistent with the allowed region reported by the LSND 
experiment \cite{lsnd}. 

\begin{acknowledgments}
We acknowledge the support of Fermilab, the Department of Energy,
and the National Science Foundation, and
we acknowledge Los Alamos National Laboratory for LDRD funding. 
\end{acknowledgments}


\bibliography{prl}

\end{document}